\documentclass[twocolumn]{aastex62}
\usepackage{natbib}
\usepackage{amsmath}
\begin{document}
\newcommand{\Qs}{Q_{\ast}'}
\newcommand{\logQs}{\log_{10}{Q_{\ast}'}}
\title{Hot Jupiters Are Destroyed by Tides While Their Host Stars Are on the Main Sequence}
\author[0000-0002-7993-4214]{Jacob H. Hamer}
\affiliation{Department of Physics and Astronomy, Johns Hopkins University, 3400 N. Charles Street, Baltimore, MD 21218, USA}\correspondingauthor{Jacob H. Hamer}\email{jhamer3@jhu.edu}

\author[0000-0001-5761-6779]{Kevin C. Schlaufman}

\affiliation{Department of Physics and Astronomy, Johns Hopkins University, 3400 N. Charles Street, Baltimore, MD 21218, USA}
\email{kschlaufman@jhu.edu}

\received{2019 June 28} \revised{2019 July 31} \accepted{2019 August 14} \submitjournal{\aj} \published{2019 October 21}
\begin{abstract}

\noindent
While cooler giant planets are often observed with nonzero eccentricities, the short-period circular orbits of hot Jupiters suggest that they lose orbital energy and angular momentum due to tidal interactions with their host stars.  However, orbital decay has never been unambiguously observed.  We use data from Gaia Data Release 2 to show that hot Jupiter host stars have a smaller Galactic velocity dispersion than a similar population of stars without hot Jupiters.  Since Galactic velocity dispersion is correlated with age, this observation implies that the population of hot Jupiter host stars is on average younger than the field population.  The best explanation for this inference is that tidal interactions cause hot Jupiters to inspiral while their host stars are on the main sequence.  This observation requires that the typical modified stellar tidal quality factor $\Qs$ for solar-type stars is in the range $\logQs \lesssim 7$.

\end{abstract}


\section{Introduction}
The apparent longevity of hot Jupiters is one of the original problems raised by the discovery of 51 Peg b \citep[e.g.,][]{Mayor1995,Rasio1996}.  When compared with the wide range of eccentricities observed in longer-period giant planet systems, the circular orbits of hot Jupiters show that tidal interactions are important at short periods \citep[e.g.,][]{Rasio1996b,Jackson2008}.  While most hot Jupiters are formally unstable to tidal decay \citep{Levrard2009}, the inspiral timescales could be longer than the main sequence lifetimes of their host stars.  As a consequence, it is uncertain whether or not hot Jupiters survive the main sequence. 

Many groups have searched for direct evidence of tidal decay in individual systems as a secularly decreasing period.  However, no studies have found unambiguous evidence of tidal inspiral \citep[e.g.,][]{Maciejewski2016,Patra2017,Wilkins2017,Maciejewski2018, Bouma2019}.  In most hot Jupiter systems, no departure from a linear ephemeris has been observed (e.g., HAT-P-23, KELT-1, KELT-16, WASP-28, WASP-33 and WASP-103).  In this case, only weak upper limits on the dissipation efficiencies can be derived.  On the other hand, while departures from linear ephemerides have been observed in the WASP-12 and WASP-4 systems, explanations other than a secularly decreasing period due to tidal dissipation have not been conclusively ruled out.

Other groups have used the ensemble properties of hot Jupiter systems to search for indirect evidence of tidal decay.  For example, it has been argued that the semimajor axis distribution of close-in giant planets shows evidence of sculpting by tidal decay \citep[e.g.,][]{Jackson2009}.  In addition, in systems in which hot Jupiters experience orbital decay, the angular momenta of their orbits should be transferred to the rotational angular momenta of their host stars.  \citet{McQuillan2013} observed a dearth of close-in giant planets around fast-rotating stars, which \citet{Teitler2014} attributed to this effect.  Moreover, \citet{Schlaufman2013} showed that hot Jupiters disappear by the time their host stars evolve off of the main sequence. 

The tidal spin-up of hot Jupiter host stars would represent an observable signal of tidal interaction. \citet{Pont2009} used rotation periods derived from rotational broadening and photometric variability to show that tidal spin-up can explain the excess rotation of some stars hosting hot Jupiters.  Excess rotation in transiting planet systems was also found by \citet{Schlaufman2010} in a search for evidence of spin-orbit misalignment. Similarly, \citet{Maxted2015} showed that at least in some systems with transiting planets, tidal spin-up can explain the observation that gyrochronological age estimates are systematically younger than isochronal age estimates.

Despite all of this progress, the effort to understand the physics of tidal dissipation in hot Jupiter systems has been held back by the difficulty of estimating robust ages for hot Jupiter host stars.  Indeed, the estimation of a precise age for an isolated field star is one of the most challenging inferences in astronomy \citep[e.g.,][]{Soderblom2010}.  While calculating accurate and precise absolute ages is still a challenge, important progress could be made with precise relative ages.  Because it is well established that the Galactic velocity dispersion of a stellar population is correlated with its age \citep[e.g.,][]{Binney2000}, comparisons of velocity dispersions reveal relative ages.  If tidal dissipation is efficient and hot Jupiters are tidally destroyed while their host stars are on the main sequence, then hot Jupiter hosts must be younger than similar field stars.  Alternatively, if tidal dissipation is inefficient and hot Jupiters survive until their host stars become subgiants, then hot Jupiter hosts and field stars should have similar age distributions.

For the first time, Gaia Data Release 2 (Gaia-DR2) \citep[][]{Lindegren2018, GaiaMission, GaiaDR2HR, GaiaDR2} has provided the necessary data to compare the ages of hot Jupiter and non-hot Jupiter host stars.  In this paper, we compare the Galactic velocity dispersion of main sequence hot Jupiter host stars to that of a matched sample of field main sequence stars.  We find that hot Jupiter hosts are kinematically colder and therefore younger than the matched sample of field stars.  For that reason, hot Jupiter hosts are a preferentially young population, and tidal dissipation is therefore efficient enough to destroy hot Jupiters while their hosts are on the main sequence. We find that only values of the modified stellar tidal quality factor $\logQs \lesssim 7$ are consistent with the data.  This is the first unambiguous evidence of the inspiral of hot Jupiters due to tidal dissipation during the main sequence lifetimes of their hosts.  This paper is organized as follows. In Section 2, we describe the creation of both our hot Jupiter host and field star samples.  In Section 3, we describe our methods to make a robust comparison between the velocity dispersions of the hot Jupiter host and field star samples.  In Section 4, we derive a limit on $\Qs$ and compare our constraint to previous measurements.  We conclude in Section 5.

\section{Data}

We obtain our list of hot Jupiter hosts from the NASA Exoplanet Archive.  We select hot Jupiters according to the fiducial definition of \citet{Wright2012}: planets with orbital period $P~<~10$ days and minimum mass $M_{p} \sin{i}~>~0.1~M_{\mathrm{Jup}}$.  We query SIMBAD for each hot Jupiter host's Gaia DR2 designation.  For objects without a radial velocity from the NASA Exoplanet Archive, we use the Gaia-DR2 radial velocity \citep{GaiaDR2RV}.

We construct our sample of field stars without hot Jupiters by querying Gaia DR2 for objects with the full five-parameter astrometric solution and a radial velocity.  While there may be undetected hot Jupiters around stars in our field sample, the low occurrence of hot Jupiters limits this possible contamination to about 1\% \citep[e.g.,][]{Wright2012, Santerne2016, Zhou2019}.  To ensure that both of our samples have reliable astrometry and radial velocities, we impose the quality cuts discussed in the Appendix. We obtain the distances to our hot Jupiter host stars and field stars from \citet{BailerJones2018}.  We use these distances when deriving absolute magnitudes, calculating extinction and reddening, and computing space velocities.  Because of the quality cuts, the median change in distance between inverted parallaxes and using the \citet{BailerJones2018} distances is less than 1\%.

\section{Analysis}
\label{section3}

\begin{figure}
    \centering
    \plotone{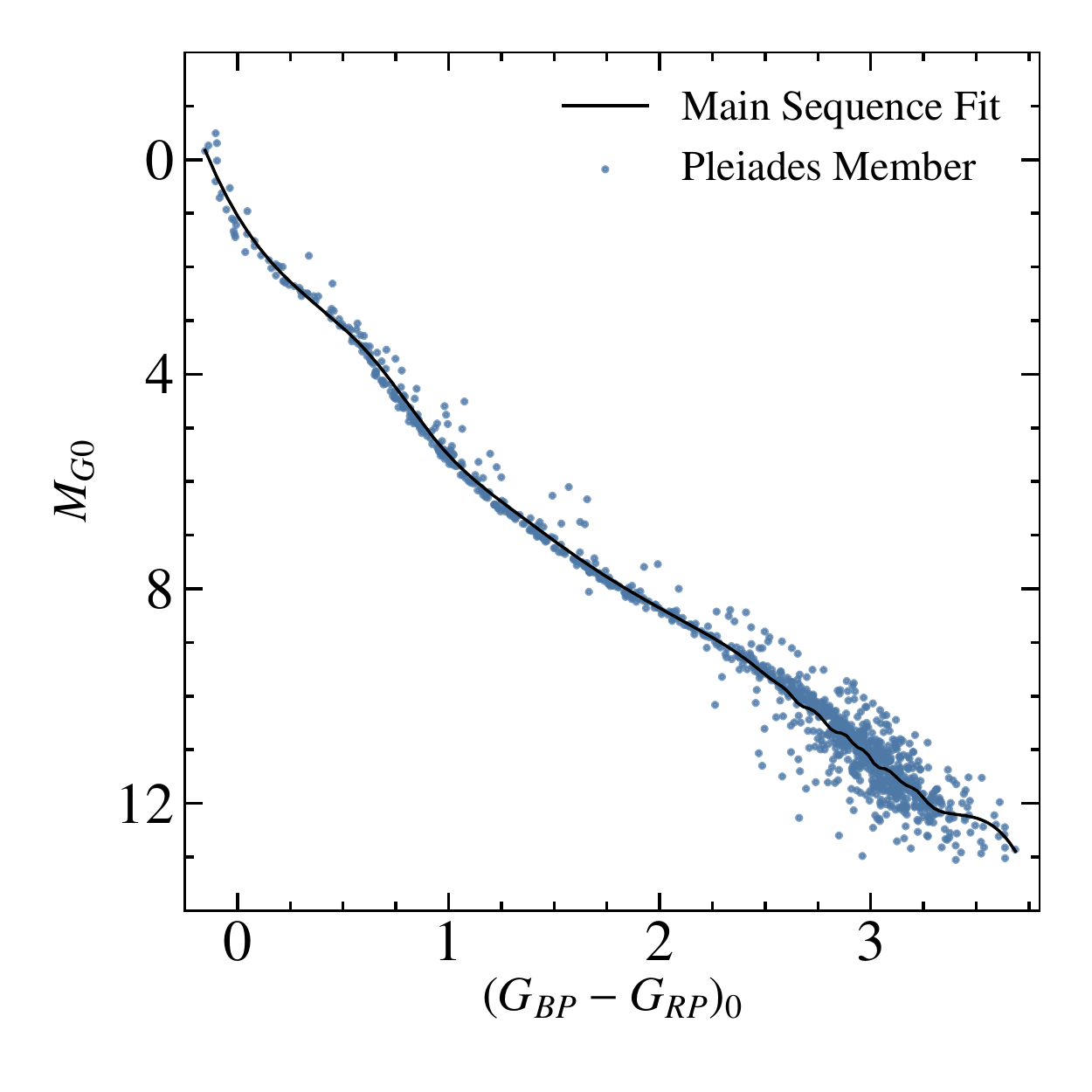}
    \caption{Members of the Pleiades in Gaia Data Release 2 from \citet{GaiaDR2HR} used to construct a polynomial fit to the zero-age main-sequence.  We first fit a smoothing spline to these data and then fit an eighth-order polynomial to the densely sampled points on the spline (shown here in black).  We use the spline fit to the main sequence to exclude all stars more than one $G$ magnitude above the main sequence from both the hot Jupiter host and control samples.}
    \label{fig:Figure 1}
\end{figure}

After selecting stars for our hot Jupiter host and field star samples, we need to reject evolved stars.  To do so, we model a solar metallicity zero-age main sequence sample using Gaia-DR2-confirmed Pleiades members \citep{GaiaDR2HR}.  We apply quality cuts to the Pleiades sample that are similar to those that we use in the construction of our field star sample.  After the application of the quality cuts, we also remove the known white dwarf EGGR 25 (Gaia-DR2 66697547870378368).  We convert from the apparent Gaia DR2 $G$ magnitude to an absolute $G$-band magnitude $M_{G}$ using the optimized parallax values for each individual member from \citet{GaiaDR2HR}.  We correct for extinction and reddening using the single value of extinction $E(B-V)=0.045$ given in Table 2 of \citet{GaiaDR2HR} and the mean extinction coefficients from \citet{Casagrande2018}.  We then fit a smoothing spline to the Pleiades data in the $(G_{\mathrm{BP}}-G_{\mathrm{RP}})_{0}$--$M_{G0}$ plane and next fit an eighth-order polynomial to the densely sampled smoothing spline.  Sorted by increasing order, the coefficients of the polynomial are $a_i=($1.07857, 6.23258, $-$10.85944, 21.65561, $-$20.27879, 9.78665, $-$2.52543, 0.33004, $-$0.0170556$)$.  We plot in Figure \ref{fig:Figure 1} both the stars used to construct the fit and the smoothing spline itself. 

To accurately limit our sample to the main sequence, we also need to account for reddening and extinction.  We calculate individual line-of-sight reddening values for each star by interpolating the three-dimensional reddening map from \citet{Capitanio2017}.  For each star, we integrate the interpolated grid along the line of sight to calculate a total $E$($B-V$) reddening.  We convert $E$($B-V$) to Gaia reddening $E$($G_{\mathrm{BP}}-G_{\mathrm{RP}}$) and extinction $A_{G}$, using the mean extinction coefficients from \citet{Casagrande2018}.

\begin{figure}
    \centering
    \plotone{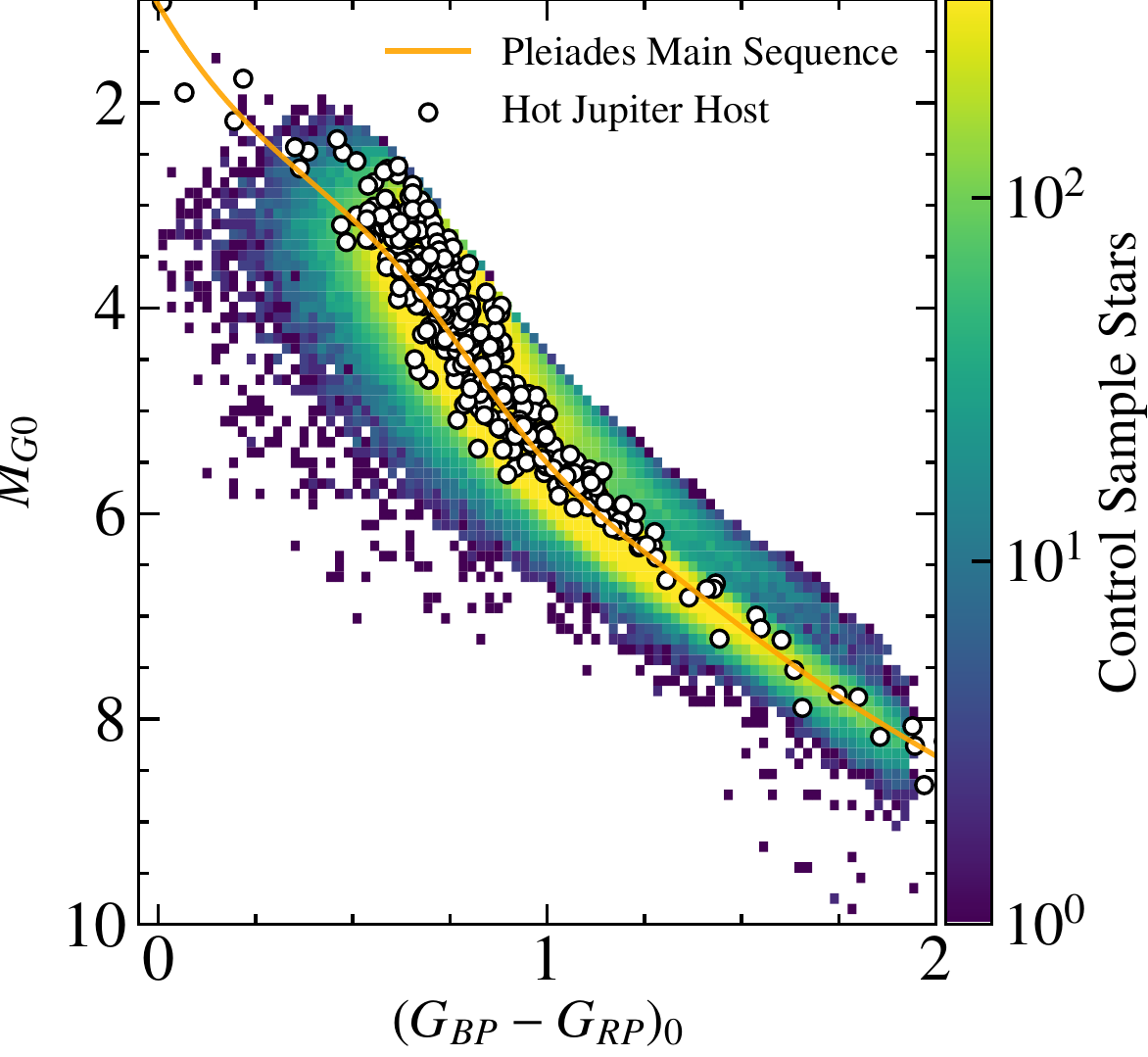}
    \caption{Hot Jupiter host and field star samples.  We plot hot Jupiter hosts as black points and the density of stars in the field star sample as the background color map.  We indicate the main sequence spline fit used to remove evolved stars as the orange line.}
    \label{fig:Figure 2}
\end{figure}

After we account for extinction and reddening in both the hot Jupiter host and field star samples, we limit both samples to objects that lie less than one magnitude above the Pleiades main sequence relation.   This results in a sample of 338 main sequence hot Jupiter host stars and 385,036 main sequence field stars.  We show the resulting hot Jupiter host and field star samples in Figure \ref{fig:Figure 2}. We provide in Table~\ref{tab:tab1} the final sample of main sequence hot Jupiter hosts that pass the astrometric quality cuts. 

\begin{deluxetable}{cc}
\tablecaption{Main Sequence hot Jupiter Hosts with Good Astrometry}
\tablenum{1}
\tablehead{\colhead{Exoplanet Archive Name} & \colhead{SIMBAD Name}}
    \startdata
         WASP-136 &  WASP-136 \\
         KELT-1 &  KELT-1 \\
         HATS-34 &  HATS-34 \\
         WASP-96 &  WASP-96 \\
         WASP-44 &  WASP-44 \\
         WASP-32 &  WASP-32 \\
         WASP-158 &  TYC 5264-1048-1 \\
         Qatar-4 &  Qatar 4 \\
         WASP-1 &  WASP-1 \\
         WASP-45 &  WASP-45 \\
    \enddata
    \tablecomments{Table 1 is ordered by right ascension and is published in its entirety in the machine-readable format.}
    \label{tab:tab1}
\end{deluxetable}

\begin{figure*}
    \centering
    \plotone{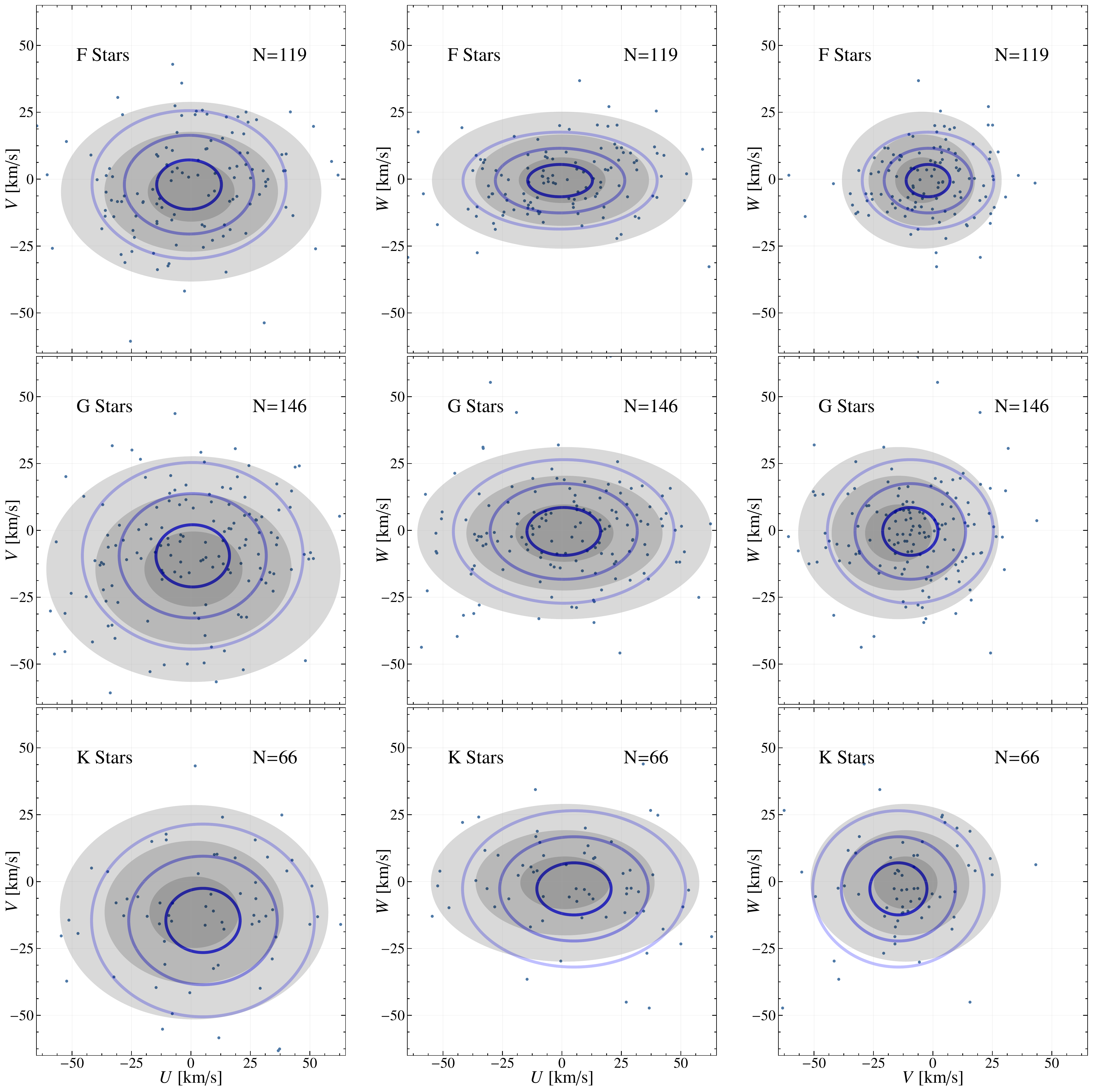}
    \caption{Galactic velocity dispersions of hot Jupiter host and field star samples.  From smallest to largest, the gray ellipses are the 1-, 2-, and 3$\sigma$ regions, and each ellipse is centered at the mean velocity of the field star sample.  The blue contours represent the same data for the hot Jupiter host star sample.  We plot individual hot Jupiter hosts as blue points.  The number in each panel represents the number of hot Jupiter hosts of the given spectral type.  The hot Jupiter hosts appear to have colder kinematics, suggesting that they are on average younger than the field stars.}
    \label{fig:Figure 3}
\end{figure*}

We use \texttt{pyia}\footnote{\url{https://zenodo.org/record/1275923\#.XKdmXEMpDCI}} to calculate the Galactic $UVW$ space velocities of the stars with respect to the local standard of rest \citep{PriceWhelan2018}.  On the main sequence, more massive blue stars live short lives, whereas less massive red stars live long lives.  Therefore, we divide the sample into F, G, and K dwarfs based on their $(G_{BP}-G_{RP})_{0}$ color to ensure that any differences in velocity dispersion are due to differences in relative population age.  We calculate the $(G_{BP}-G_{RP})_{0}$ colors that correspond to F, G, and K stars by using the $T_{\mathrm{eff}}$ derived in Gaia-DR2 for stars within the Pleiades.  We use data from \citet{Mamajek2013} to define F, G, and K spectral types by their effective temperatures: F stars with $6000$ K $< T_{\mathrm{eff}} < 7500$ K have $0.438 < (G_{BP}-G_{RP})_{0} < 0.742$, G stars with $5340$ K $< T_{\mathrm{eff}} < 6000$ K have $0.742 < (G_{BP}-G_{RP})_{0} < 1.00$, and K stars with $3940$ K $< T_{\mathrm{eff}} < 5340$ K have $1.00 < (G_{BP}-G_{RP})_{0} < 2.02$.  We then model the $UVW$ velocity distributions as Gaussians and represent these Gaussian distributions with ellipses in Figure~\ref{fig:Figure 3}.  We find that hot Jupiter host stars have a colder velocity dispersion than the field star sample in each subsample, suggesting that they are a younger population. 

To verify that our calculation in Figure~\ref{fig:Figure 3} is not affected by any other characteristics correlated with hot Jupiter occurrence, we perform a Monte Carlo simulation.  Since hot Jupiter occurrence is correlated with stellar metallicity \citep[e.g.,][]{Santos2004, Fischer2005}, it is important to ensure a matched proportion of thin/thick disk stars in the control sample.  Metal-poor stars are more likely to be members of the thick disk, whereas metal-rich stars are more likely to be members of the thin disk.  Because height above or below the Milky Way midplane $z$ is related to age, metallicity, and/or thin/thick disk membership, matching on the easily calculated $z$ distribution accounts for possible correlations of hot Jupiter occurrence with age, metallicity, and thin/disk membership.  On each iteration, we select a subsample of the field star sample that is matched in its $z$ distribution to the hot Jupiter host sample.

\begin{figure*}
    \centering
    \plotone{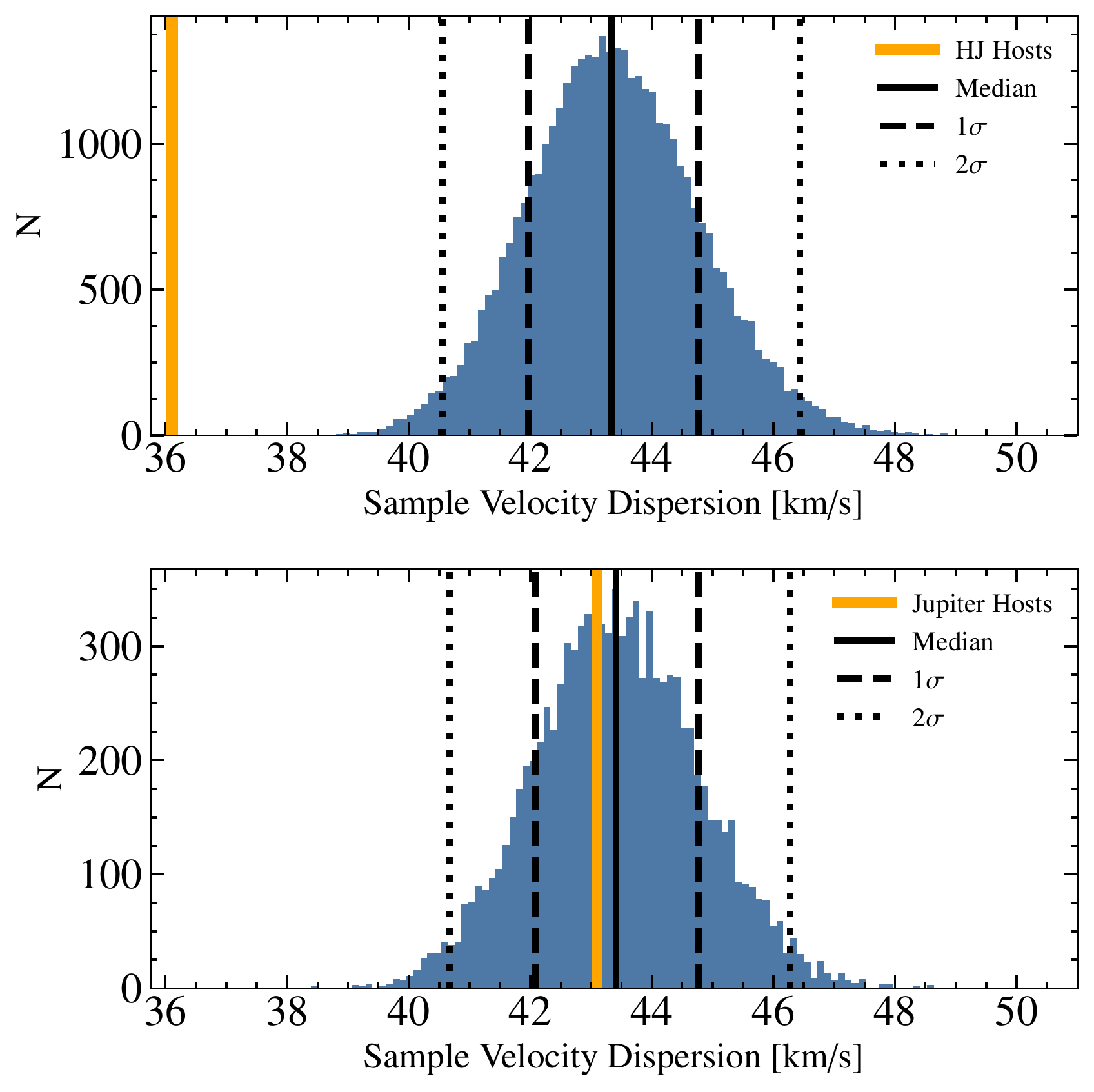}
    \caption{Velocity dispersion distribution of the matched control samples (blue histograms) compared to the velocity dispersion of our exoplanet host samples (orange vertical lines).  The black vertical lines show the (2nd, 16th, 50th, 84th, and 98th) percentiles of the Monte Carlo samples.  Top: the velocity dispersion of the hot Jupiter host sample.  Less than one in 40,000 Monte Carlo samples has a velocity dispersion as small as that of the hot Jupiter host sample.  In other words, the hot Jupiter hosts are significantly kinematically colder than similar non-hot Jupiter host stars.  Bottom: the velocity dispersion of the longer-period giant planet host sample.  The longer-period giant planet hosts have a velocity dispersion that is fully consistent with that of the field star sample.  If the smaller velocity dispersion of hot Jupiter hosts in the top panel was unrelated to tidal dissipation, then this sample of longer-period planets would exhibit a similarly small velocity dispersion.  The only explanation for this observation is that hot Jupiter hosts are on average a younger population than non-hot Jupiter host stars.  The implication is that hot Jupiter orbits decay due to tidal dissipation and that these planets are destroyed before their hosts evolve off of the main sequence.}
    \label{fig:Figure 4}
\end{figure*}

Since it is still important to maintain similar color distributions in the hot Jupiter host and $z$-matched field star sample, on each iteration we construct a $z$- and color-matched control sample by selecting 338 stars from the $z$-matched sample such that every hot Jupiter host is accounted for by a star in the control sample within 0.025 mag in $(G_{BP}-G_{RP})_{0}$.  The result is a single Monte Carlo control sample that is matched as well as possible to the hot Jupiter host sample.  For each of these Monte Carlo iterations, we calculate the mean $UVW$ velocity and then calculate the $UVW$ velocity dispersion
\begin{equation}
    \frac{1}{N}\sum \left[(U_i-\overline{U})^2+(V_i-\overline{V})^2+(W_i-\overline{W})^2\right]^{1/2}.
\end{equation}
We plot the result of this Monte Carlo simulation in the top panel of Figure \ref{fig:Figure 4}.

We find that the sample of hot Jupiter host stars has a velocity dispersion significantly colder than that of the matched Monte Carlo samples of field stars.  The probability that the hot Jupiter host velocities were drawn from the same parent distribution as the field star sample is less than one in 40,000 (or more than 4$\sigma$).  We are limited in terms of the number of Monte Carlo iterations that can be computed, so the significance could exceed 4$\sigma$.  The sample of hot Jupiter hosts must therefore be systematically younger than a matched field star population.  We argue that the only tenable explanation for this observation is that hot Jupiters lose orbital energy due to tidal dissipation and inspiral to their destruction while their host stars are on the main sequence.

We verify our interpretation of this result by making a similar comparison between the hosts of longer-period giant planets and an equivalent field star sample.  If the cold velocity dispersion of the hot Jupiter host sample is a result of the tidal destruction of hot Jupiters while their host stars are on the main sequence, then main sequence stars that host longer-period giant planets should have kinematics consistent with those of the field star sample.  On the other hand, if the hot Jupiter host sample was biased toward young stars as a result of the combination of a possible age--metallicity relationship and the well-known connection between host star metallicity and giant planet occurrence, then the hot Jupiter and longer-period giant planet host samples should both have cold velocity dispersions.  It should be noted that recent studies find no evidence for an age--metallicity relationship in the thin disk \citep[e.g.,][]{Casagrande2011, Bensby2014, SilvaAguirre2018}.  We construct a sample of 367 longer-period giant planet hosts with $P~>~10 $ days and $M_{p} \sin{i}~>~0.1~M_{\mathrm{Jup}}$, by use of the same process described in Section 2.  We note that the hot Jupiter and longer-period giant planet host samples are of similar size, and we use an equivalent field star sample to carry out our Monte Carlo sampling as previously described.  The hot Jupiter host star, longer-period Jupiter host star, and field star samples have similar spatial distributions, so our kinematic comparisons should not be affected by any relationships between location in the galaxy and kinematics. As we match our Monte Carlo control samples in their distribution relative to the Milky Way midplane, we control for known correlations between spatial distribution and kinematics.  We find that longer-period giant planet hosts have kinematics indistinguishable from those of the field stars.  As a result, we conclude that the difference in velocity dispersion between the hot Jupiter host stars and the matched samples of field stars is a consequence of tidal dissipation.

\section{Discussion}
The velocity dispersion of main sequence hot Jupiter host stars is significantly colder than that of an equivalent sample of field stars.  Since there is no reason to believe that hot Jupiter formation is favored in a population with low velocity dispersion \citep[e.g.,][]{McTier2019}, this indicates that hot Jupiter hosts are on average younger than similar field stars.  The implication is that hot Jupiters are a transient population that experience rapid inspiral and tidal disruption during the main sequence lifetimes of their host stars.  This is the first unambiguous evidence of the destruction of hot Jupiters during the main sequence lifetimes of their host stars.

Although we argue that our observation is definitive, other authors have reached similar conclusions based on other methods.  \citet{Jackson2009} argued that the semimajor axis distribution of close-in giant planets showed evidence of sculpting by tides, but that approach required assumptions about the unknown initial period and mass distribution of hot Jupiters.  \citet{McQuillan2013} observed a dearth of close-in planets around fast-rotating stars in the  Q1-Q6 {\it Kepler} Object of Interest list \citep{Batalha2013}, which \citet{Teitler2014} attributed to stellar tidal spin-up by inspiraling planets.  In support of the robustness of their observation, \citet{McQuillan2013} argued that the infrequency of close-in planets around fast rotators cannot be linked to the increased photometric variability typical of these stars because long-period planets that were potentially more difficult to detect were observed.  This argument does not account for the fact that the product of planet occurrence and transit probability goes up in this range in orbital period \citep[e.g.,][]{Fressin2013, Hsu2019}.  Additionally, many fast-rotating stars in the Kepler field will be in short-period nontransiting binary systems that cannot host short-period planets.  Furthermore, the short periods of close-in planets indicate that planet and activity signals are not well separated in frequency, hindering the detection of close-in planets around fast-rotating stars.  Finally, fast-rotating stars, either young or massive, will have large radii, which make it difficult to discover transiting planets.

Having established that hot Jupiters inspiral while their hosts are on the main sequence, our goal now is to more precisely quantify this effect.  The semimajor axis evolution of a circularized orbit is driven by dissipation inside the primary \citep[e.g.,][]{Rasio1996, Matsumura2010}.  The dissipation efficiency inside of the primary is usually characterized by the modified stellar tidal quality factor $\Qs$, a parameter that is approximately the reciprocal of the ratio of the amount of energy stored in the tidal disturbance to the amount of energy dissipated over one cycle.  Tidal dissipation is highly nonlinear and depends on star and planet mass, the system orbital period, and detailed models of the interior structures of both star and planet.  Therefore, there is significant uncertainty regarding the efficiency of tidal dissipation, despite decades of theoretical work.

Different models for these detailed processes indicate that $\Qs$ can vary by three orders of magnitude.  There are two components to the tidal disturbance: dynamical tides and equilibrium tides.  Dynamical tides refer to both the excitation of inertial waves in convective zones restored by the Coriolis force and to internal gravity waves in the stably stratified radiative zone restored by gravity.  The dissipation of inertial waves can produce $\logQs \sim 6$, but they are not expected to operate in most hot Jupiter systems because their tidal forcing frequencies usually do not satisfy $|\omega|<2\Omega_{\ast}$, where $\omega$ is the tidal forcing frequency and $\Omega_{\ast}$ is the stellar spin frequency \citep[e.g.,][]{Ogilvie2004, Ogilvie2009}.  Internal gravity waves can produce $\logQs\sim 5$ for hot Jupiters in the nonlinear regime with $M_{\mathrm{p}}\gtrsim 3~M_{\mathrm{Jup}}$ \citep[e.g.,][]{Barker2010}, but these planets are rare in our sample and in general \citep[e.g.,][]{Cumming2008}.  Simulations predict $\Qs$ in the range $5\lesssim\logQs\lesssim6$ for planets with $M_{\mathrm{p}} \gtrsim 0.5~M_{\mathrm{Jup}}$ and $ P\lesssim 2\ \text{days}$ in the weakly nonlinear regime, but only 10\% of our sample is in this part of parameter space \citep[e.g.,][]{Essick2016}.  The expectation is therefore that most of the tidal dissipation in our hot Jupiter host sample is likely dominated by the equilibrium tide.  The term equilibrium tides refers to tidally induced large scale velocity flows in a body.  Most hot Jupiter hosts have convective envelopes, so in this case the tidal energy is dissipated by convective turbulence \citep[e.g.,][]{Zahn1977, Hut1981,Eggleton1998}.  \citet{Penev2011} numerically modeled this process and found that $8\lesssim\logQs\lesssim9$.

Given the complexity of tidal dissipation, observational constraints on its efficiency are important.  On an observational basis, we determine the range of $\Qs$ that would cause the inspiral of hot Jupiters in our sample during the main sequence lifetimes of their host stars.  We focus on hot Jupiter systems with homogeneously determined stellar parameters.  We use both the SWEET-Cat \citep{Santos2013, Andreasen2017, Sousa2018} catalog and the catalogs of \citet{Brewer2016} and \citet{Brewer2018}.

\begin{figure*}
    \centering
    \plottwo{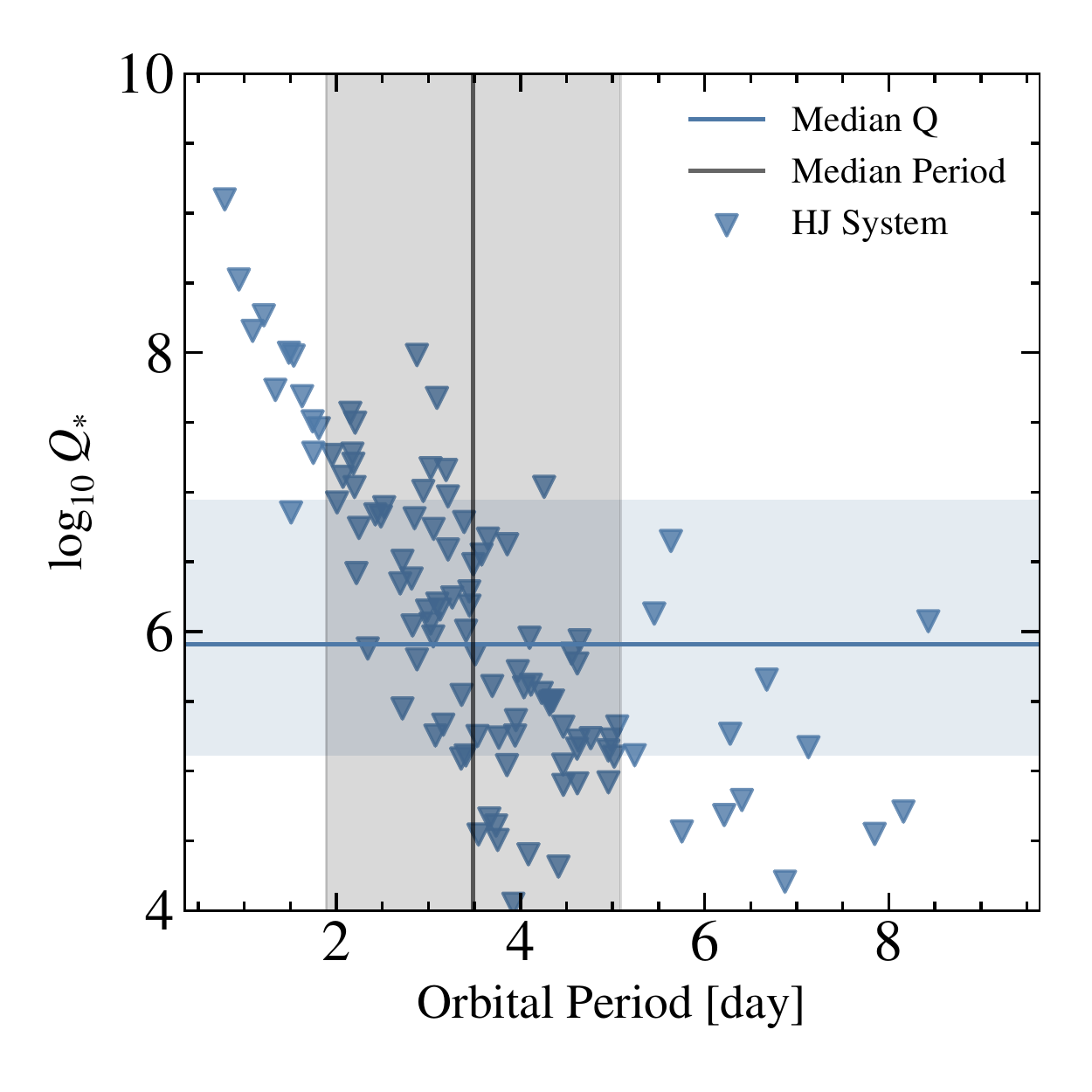}{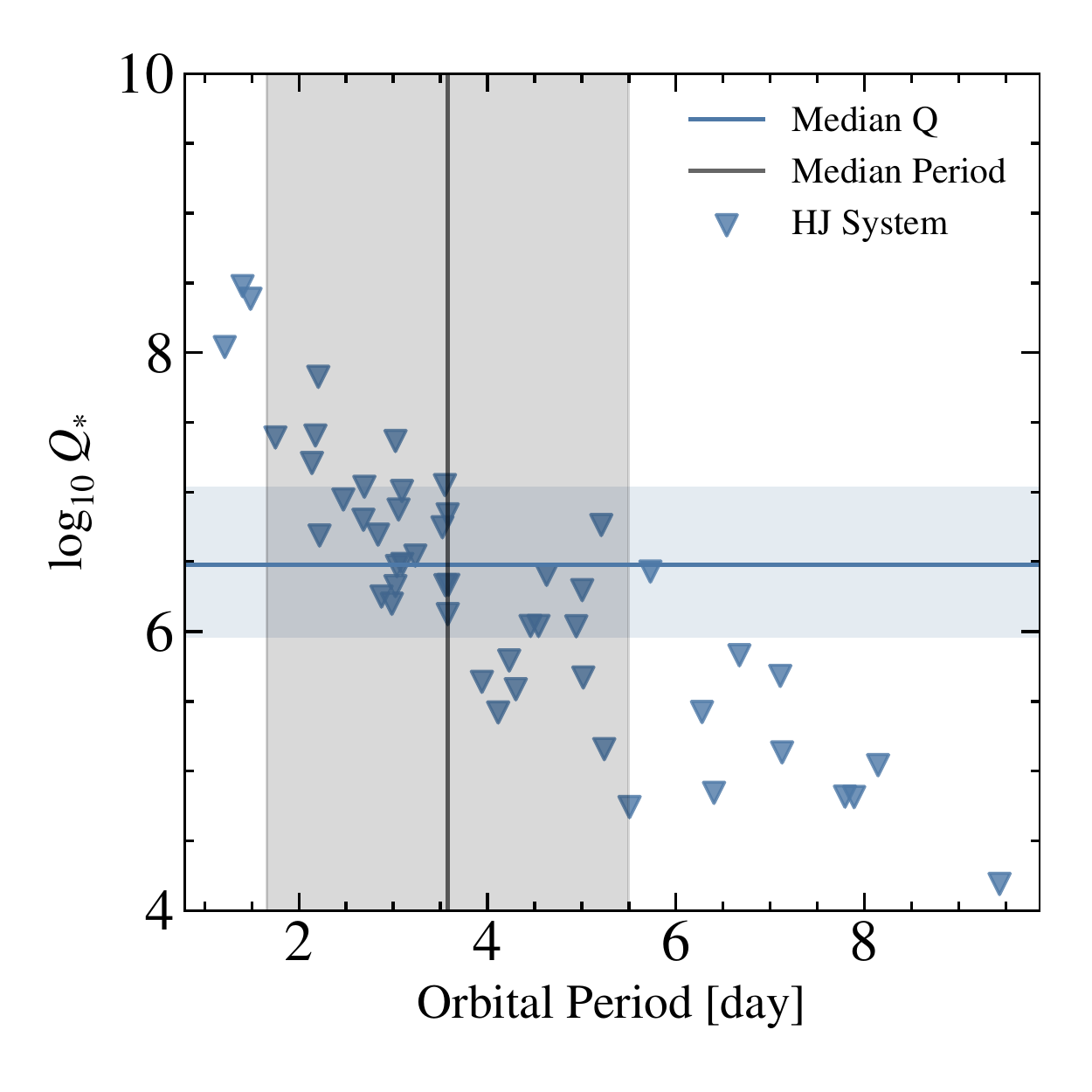}
    \caption{Maximum $\Qs$ required for tidal decay during the main sequence.  We use the formalism presented in \citet{Lai2012} and homogeneously derived spectroscopic stellar parameters from the literature to derive the limit on $\Qs$.  The vertical line shows the median period of the Jupiter sample, whereas the gray rectangle spans the 16th to 84th percentiles of the sample period distribution (approximately 2 days $\lesssim P \lesssim$ 5 days).  We calculate the maximum $\Qs$ required for each system, and within this period range where our velocity dispersion analysis applies, we calculate the (16,50,84) percentiles.  For those systems in the gray rectangle, the horizontal line is the median $\Qs$ and the blue rectangle spans the 16th to 84th percentiles of the inferred $\Qs$ distribution.  Left: using spectroscopic stellar parameters from the SWEET-Cat catalog of \citet{Santos2013}, we find $\logQs< 5.95^{+0.98}_{-0.83}$.  Right: using spectroscopic stellar parameters from \citet{Brewer2016} and \citet{Brewer2018}, we find $\logQs< 6.48^{+0.57}_{-0.52}$.  The two values are statistically indistinguishable.}
    \label{fig:Q_UL}
\end{figure*}

We first need to estimate the main sequence lifetime of each hot Jupiter host star in our sample, and we do so using the scaling relation 
\begin{equation}\label{ms_scale}
    \frac{t_{\mathrm{MS,}\ast}}{t_{\mathrm{MS,}\odot}}=\left(\frac{M_{\ast}}{M_{\odot}}\right)^{-2.5}.
\end{equation} 
To estimate the radius of each star $R_{\ast}$ in the SWEET-Cat subsample, we use the polynomial fit from \citet{Torres2010}:
\begin{equation}\label{r_scale}
\begin{split}
    \log{R_{\ast}} = b_1 + b_2X + b_3X^2+b_4X^3+b_5(\log{g})^2+\\b_6(\log{g})^3+b_7[\mathrm{Fe/H}],
\end{split}
\end{equation}
where $X=\log T_{\mathrm{eff}}-4.1$ and $b_i=($2.4427, 0.6679, 0.1771, 0.705, $-$0.21415, 0.02306, 0.04173$)$. We use the \citet{Brewer2016} and \citet{Brewer2018} isochrone-derived values for $R_{\ast}$.  It has been shown that a reasonable approximation for the inspiral time $t_{\mathrm{in}}$ is $t_{\mathrm{in}}=(2/13)t_{\mathrm{a}}$, where $t_{\mathrm{a}} = a/\dot{a}$ is defined as follows \citep[e.g.,][]{Barker2009,Lai2012}:
\begin{equation}\label{t_a}
    t_{\mathrm{in}} = \frac{2}{13} t_{\mathrm{a}} = \frac{2}{13} \frac{2\Qs}{9}\frac{M_*}{M_{\mathrm{p}}}\left(\frac{a}{R_*}\right)^5\frac{P}{2\pi}.
\end{equation} 
Finally, we solve Equation (\ref{t_a}) for $\Qs$, assuming $t_{\mathrm{in}} = t_{\mathrm{MS}}$ to obtain an upper limit on $\Qs$ for each system
\begin{equation}\label{Q_ul}
    \Qs < t_{\mathrm{MS}}\frac{117}{4}\frac{M_{\mathrm{p}}}{M_{\ast}}\left(\frac{R_{\ast}}{a}\right)^5\frac{2\pi}{P}.
\end{equation}
Assuming the full main sequence lifetime is conservative.  It is a reasonable assumption, however, because a star's radius grows on the main sequence, and the tidal decay timescale is a strong function of that radius.  We therefore argue that decay will most likely occur close to the end of a star's main sequence lifetime.  We plot the results of this calculation in Figure~\ref{fig:Q_UL}.  Because we use a population-level approach, we can only provide constraints based on the typical system within our sample.  We estimate $\Qs$ in our typical hot Jupiter system by calculating the median $\Qs$ among the systems with periods that fall within the 16th and 84th period percentiles.  Because our sample is dominated by transit-detected hot Jupiters, our estimate is only valid within the period range 2 days $ \lesssim P \lesssim $ 5 days.  We find that the destruction of hot Jupiters before the end of their host's main sequence lifetime requires $\logQs< 5.95^{+0.98}_{-0.83}$ when using spectroscopic stellar parameters from SWEET-Cat. When using spectroscopic stellar parameters from \citet{Brewer2016} and \citet{Brewer2018}, we find $\logQs< 6.48^{+0.57}_{-0.52}$.

One may be concerned that our result is not evidence of the tidal destruction of hot Jupiters but, rather, a consequence of a bias toward observing systems early in their tidal evolution when orbital decay is slowest. We investigate whether this is so by comparing to the field sample the velocity dispersions of hot Jupiter host star subsamples partitioned by relative inspiral time. Assuming $\logQs=6$, we calculate the inspiral time for every hot Jupiter system in the SWEET-Cat catalog by use of Equation~\ref{t_a}. We divide this sample in half according to inspiral time, and we repeat the Monte Carlo experiment described in Section~\ref{section3} for each half. We find that the velocity dispersion of both subsamples is smaller than that of the field, with similar statistical significances noted. This supports our interpretation of the colder kinematics of hot Jupiter hosts serving as evidence for the tidal destruction of hot Jupiters.

Our inferred value of $\Qs$ agrees with estimates produced by searches for secularly decreasing orbital periods in the most extreme hot Jupiter systems.  Most of these analyses have found period derivatives $\dot{P}$ consistent with zero, producing lower limits on the order of $\logQs \gtrsim 5-6$ \citep[e.g.,][]{Hoyer2016,Jiang2016,Wilkins2017,Maciejewski2018}.  \citet{Maciejewski2018} and \citet{Bouma2019} have observed departures from a linear ephemeris in the WASP-12 and WASP-4 systems.  If associated with tidal decay, these observations require $\Qs=(1.82\pm0.32)\times10^5$ for WASP-12 and $\Qs=(2.9\pm0.3)\times10^4$ for WASP-4.  These $\Qs$ are consistent with our upper limits.

\begin{deluxetable}{cc}
\tablecaption{Empirical Constraints on $\Qs$}
\tablenum{2}
\tablehead{\colhead{Literature Source} & \colhead{$\Qs$}}
    \startdata
        \citet{Meibom2005} & $\logQs\sim5.5$\\
        \citet{Jackson2008} & $\logQs\sim5.5$ \\
        \citet{Penev2012} & $\logQs>7$\\
        \citet{Husnoo2012} & $\logQs\sim6.5$ \\
        \citet{Hansen2012} & $7<\logQs<9$ \\
        \citet{Milliman2014} & $\logQs\sim5.5$\\
        \citet{Bonomo2017} & $\logQs\gtrsim 6-7$\\
        \citet{Penev2018} & $5 \lesssim \logQs \lesssim 7$\\
        \citet{CollierCameron2018} & $\logQs=8.3\pm0.14$\\
        \citet{LabadieBartz2019} & $\logQs\sim 6.2$\\
    \enddata
    \caption{Empirical constraints on the value of $\Qs$ from the literature.}
    \label{tab:empirical_Q}
\end{deluxetable}
 
Our upper limit on $\Qs$ is in agreement with most of the existing indirect, empirical estimates of $\Qs$ as well.  In Table~\ref{tab:empirical_Q} we present some constraints on $\Qs$ that currently exist in the literature.  The constraints on $\Qs$ in Table~\ref{tab:empirical_Q} relied on precise ages for individual hot Jupiter hosts \citep{Jackson2008, Bonomo2017, Penev2018, LabadieBartz2019}, were derived from stellar binaries \citep{Meibom2005, Milliman2014}, or made assumptions about the initial period distribution of hot Jupiters \citep{CollierCameron2018}. Our result makes no such assumptions.  Specifically, \citet{CollierCameron2018} note that their estimate relies on the assumption of a uniform distribution in $\log{P}$, as was inferred from early Doppler surveys \citep[e.g.,][]{Udry2007}.  However, a more recent analysis of confirmed transiting giant planets in the {\it Kepler} field does not support a uniform distribution in $\log{P}$ \citep{Santerne2016}.  Because transit surveys can search orders of magnitude more stars for hot Jupiters than can Doppler surveys, the period distribution inferred from long-duration transit surveys is a better approximation of the true period distribution.  Similarly, analyses that rely on precise host star ages are hindered by the fact that isochronal ages for main sequence stars typically have uncertainties larger than 20\% \citep[e.g.,][]{Soderblom2010}.  In addition, estimates of $\Qs$ derived from stellar binaries may be biased because there is evidence that the value of $\Qs$ depends on the mass of the secondary \citep[e.g.,][]{Barker2010, Ogilvie2014}.  Finally, while the limits on $\Qs$ derived by \citet{Hansen2012} do not make these assumptions, the analysis relies on systems with host stars near the transition between convective and radiative envelopes that may have less efficient dissipation \citep{Penev2018}. 

We have developed a model-independent method which allows us to place constraints on the efficiency of tidal dissipation while avoiding commonly made assumptions. While there only existed sufficiently precise astrometric data for a handful of hot Jupiter hosts before Gaia-DR2, it is now available for hundreds of hosts. Our method can be applied to other questions related to tidal dissipation, all of which would traditionally rely on assumptions like those above. 

It has been theorized that the efficiency of tidal dissipation may depend on the amplitude of the tidal forcing \citep[e.g.,][]{Ogilvie2009}.  We can test if this is true by applying our method to ultra-short-period (USP) planets, whose masses are two orders of magnitude lower than those of hot Jupiters \citep[][]{SanchisOjeda2014}. If $\Qs$ is independent of secondary mass in the planet mass regime, then our observation of orbital decay in hot Jupiters requires tidal decay in USP planets as well.  If we fail to detect a similar signal in USP planet hosts, then this would be an indication that $\Qs$ does depend on period or secondary mass.  

The apparent increased incidence of significant stellar obliquities in hot Jupiter systems with a massive/hot primary may be connected to less efficient dissipation in massive stars, which have less mass in their convective envelopes \citep{Schlaufman2010, Winn2010, Albrecht2012}. Our method provides a way to investigate the efficiency of obliquity damping in hot Jupiter systems with massive/hot star primaries.  If the alignment process is less efficient in systems with massive/hot host stars, then aligned systems observed around massive/hot stars should be on average older than misaligned systems.

\section{Conclusion}
The stability of hot Jupiters against inspiral due to tidal dissipation has been a long-standing topic of discussion in the field of exoplanets. Using their Galactic velocity dispersion, we show that main sequence hot Jupiter host stars are preferentially younger than a matched sample of field stars.  We confirm that this signal is due to tidal dissipation by showing that it disappears for stars hosting longer-period giant planets.  This is unambiguous observational evidence for the tidal destruction of hot Jupiters during the main sequence lifetimes of their host stars. Our inference is independent of the unobserved initial period distribution of hot Jupiters, precise individual host star ages, and the details of the physical processes behind tidal evolution.  Our observation requires that $\logQs \lesssim 7$ in the period range 2 days $\lesssim P \lesssim$ 5 days and the mass range 0.5 $M_{\mathrm{Jup}}\lesssim M_{p} \lesssim$ 2 $M_{\mathrm{Jup}}$.

\acknowledgements
We thank Ruth Angus, Andy Casey, Kaloyan Penev, Andrew Collier-Cameron, Brian Jackson, Simon Albrecht, and Melvyn Davies for helpful comments on this paper. This research has made use of the NASA Exoplanet Archive, which is operated by the California Institute of Technology under contract with NASA under the Exoplanet Exploration Program.  This research has made use of the SIMBAD database, operated at Centre de Donn\'ees astronomiques de Strasbourg (CDS), Strasbourg, France.  This research has made use of the VizieR catalog access tool, CDS, Strasbourg, France (DOI: \href{http://doi.org/10.26093/cds/vizier}{10.26093/cds/vizier}). The original description of the VizieR service was published in A\&AS 143, 23 \citep{SIMBAD}. This research made use of Astropy,\footnote{http://www.astropy.org} a community-developed core Python package for Astronomy \citep{astropy:2013, astropy:2018}.  This project was developed in part at the 2018 NYC Gaia Sprint, hosted by the Center for Computational Astrophysics of the Flatiron Institute in New York City, New York.  This work has made use of data from the European Space Agency (ESA) mission {\it Gaia} (\url{https://www.cosmos.esa.int/gaia}), processed by the {\it Gaia} Data Processing and Analysis Consortium (DPAC, \url{https://www.cosmos.esa.int/web/gaia/dpac/consortium}).  Funding for the DPAC has been provided by national institutions, in particular the institutions participating in the {\it Gaia} Multilateral Agreement.

\software{pyia \citep{PriceWhelan2018}, pandas \citep{pandas}, Astropy \citep{astropy:2013, astropy:2018}}
\bibliography{mybib}

\begin{thebibliography}{}
\expandafter\ifx\csname natexlab\endcsname\relax\def\natexlab#1{#1}\fi

\bibitem[{{Albrecht} {et~al.}(2012){Albrecht}, {Winn}, {Johnson}, {Howard},
  {Marcy}, {Butler}, {Arriagada}, {Crane}, {Shectman}, {Thompson}, {Hirano},
  {Bakos}, \& {Hartman}}]{Albrecht2012}
{Albrecht}, S., {Winn}, J.~N., {Johnson}, J.~A., {et~al.} 2012, \apj, 757, 18

\bibitem[{{Andreasen} {et~al.}(2017){Andreasen}, {Sousa}, {Tsantaki},
  {Teixeira}, {Mortier}, {Santos}, {Su{\'a}rez-Andr{\'e}s}, {Delgado-Mena}, \&
  {Ferreira}}]{Andreasen2017}
{Andreasen}, D.~T., {Sousa}, S.~G., {Tsantaki}, M., {et~al.} 2017, \aap, 600,
  A69

\bibitem[{{Astropy Collaboration} {et~al.}(2013){Astropy Collaboration},
  {Robitaille}, {Tollerud}, {Greenfield}, {Droettboom}, {Bray}, {Aldcroft},
  {Davis}, {Ginsburg}, {Price-Whelan}, {Kerzendorf}, {Conley}, {Crighton},
  {Barbary}, {Muna}, {Ferguson}, {Grollier}, {Parikh}, {Nair}, {Unther},
  {Deil}, {Woillez}, {Conseil}, {Kramer}, {Turner}, {Singer}, {Fox}, {Weaver},
  {Zabalza}, {Edwards}, {Azalee Bostroem}, {Burke}, {Casey}, {Crawford},
  {Dencheva}, {Ely}, {Jenness}, {Labrie}, {Lim}, {Pierfederici}, {Pontzen},
  {Ptak}, {Refsdal}, {Servillat}, \& {Streicher}}]{astropy:2013}
{Astropy Collaboration}, {Robitaille}, T.~P., {Tollerud}, E.~J., {et~al.} 2013,
  \aap, 558, A33

\bibitem[{{Bailer-Jones} {et~al.}(2018){Bailer-Jones}, {Rybizki}, {Fouesneau},
  {Mantelet}, \& {Andrae}}]{BailerJones2018}
{Bailer-Jones}, C.~A.~L., {Rybizki}, J., {Fouesneau}, M., {Mantelet}, G., \&
  {Andrae}, R. 2018, \aj, 156, 58

\bibitem[{{Barker} \& {Ogilvie}(2009)}]{Barker2009}
{Barker}, A.~J., \& {Ogilvie}, G.~I. 2009, \mnras, 395, 2268

\bibitem[{{Barker} \& {Ogilvie}(2010)}]{Barker2010}
---. 2010, \mnras, 404, 1849

\bibitem[{{Batalha} {et~al.}(2013){Batalha}, {Rowe}, {Bryson}, {Barclay},
  {Burke}, {Caldwell}, {Christiansen}, {Mullally}, {Thompson}, {Brown},
  {Dupree}, {Fabrycky}, {Ford}, {Fortney}, {Gilliland}, {Isaacson}, {Latham},
  {Marcy}, {Quinn}, {Ragozzine}, {Shporer}, {Borucki}, {Ciardi}, {Gautier},
  {Haas}, {Jenkins}, {Koch}, {Lissauer}, {Rapin}, {Basri}, {Boss}, {Buchhave},
  {Carter}, {Charbonneau}, {Christensen-Dalsgaard}, {Clarke}, {Cochran},
  {Demory}, {Desert}, {Devore}, {Doyle}, {Esquerdo}, {Everett}, {Fressin},
  {Geary}, {Girouard}, {Gould}, {Hall}, {Holman}, {Howard}, {Howell},
  {Ibrahim}, {Kinemuchi}, {Kjeldsen}, {Klaus}, {Li}, {Lucas}, {Meibom},
  {Morris}, {Pr{\v s}a}, {Quintana}, {Sanderfer}, {Sasselov}, {Seader},
  {Smith}, {Steffen}, {Still}, {Stumpe}, {Tarter}, {Tenenbaum}, {Torres},
  {Twicken}, {Uddin}, {Van Cleve}, {Walkowicz}, \& {Welsh}}]{Batalha2013}
{Batalha}, N.~M., {Rowe}, J.~F., {Bryson}, S.~T., {et~al.} 2013, \apjs, 204, 24

\bibitem[{{Bensby} {et~al.}(2014){Bensby}, {Feltzing}, \& {Oey}}]{Bensby2014}
{Bensby}, T., {Feltzing}, S., \& {Oey}, M.~S. 2014, \aap, 562, A71

\bibitem[{{Binney} {et~al.}(2000){Binney}, {Dehnen}, \&
  {Bertelli}}]{Binney2000}
{Binney}, J., {Dehnen}, W., \& {Bertelli}, G. 2000, \mnras, 318, 658

\bibitem[{{Bonomo} {et~al.}(2017){Bonomo}, {Desidera}, {Benatti}, {Borsa},
  {Crespi}, {Damasso}, {Lanza}, {Sozzetti}, {Lodato}, {Marzari}, {Boccato},
  {Claudi}, {Cosentino}, {Covino}, {Gratton}, {Maggio}, {Micela}, {Molinari},
  {Pagano}, {Piotto}, {Poretti}, {Smareglia}, {Affer}, {Biazzo}, {Bignamini},
  {Esposito}, {Giacobbe}, {H{\'e}brard}, {Malavolta}, {Maldonado}, {Mancini},
  {Martinez Fiorenzano}, {Masiero}, {Nascimbeni}, {Pedani}, {Rainer}, \&
  {Scandariato}}]{Bonomo2017}
{Bonomo}, A.~S., {Desidera}, S., {Benatti}, S., {et~al.} 2017, \aap, 602, A107

\bibitem[{Bouma {et~al.}(2019)Bouma, Winn, Baxter, Bhatti, Dai, Daylan,
  D{\'{e}}sert, Hill, Kane, Stassun, Villasenor, Ricker, Vanderspek, Latham,
  Seager, Jenkins, Berta-Thompson, Col{\'{o}}n, Fausnaugh, Glidden, Guerrero,
  Rodriguez, Twicken, \& Wohler}]{Bouma2019}
Bouma, L.~G., Winn, J.~N., Baxter, C., {et~al.} 2019, The Astronomical Journal,
  157, 217

\bibitem[{Brewer \& Fischer(2018)}]{Brewer2018}
Brewer, J.~M., \& Fischer, D.~A. 2018, The Astrophysical Journal Supplement
  Series, 237, 38

\bibitem[{Brewer {et~al.}(2016)Brewer, Fischer, Valenti, \&
  Piskunov}]{Brewer2016}
Brewer, J.~M., Fischer, D.~A., Valenti, J.~A., \& Piskunov, N. 2016, The
  Astrophysical Journal Supplement Series, 225, 32

\bibitem[{{Capitanio} {et~al.}(2017){Capitanio}, {Lallement}, {Vergely},
  {Elyajouri}, \& {Monreal-Ibero}}]{Capitanio2017}
{Capitanio}, L., {Lallement}, R., {Vergely}, J.~L., {Elyajouri}, M., \&
  {Monreal-Ibero}, A. 2017, \aap, 606, A65

\bibitem[{{Casagrande} {et~al.}(2011){Casagrande}, {Sch{\"o}nrich}, {Asplund},
  {Cassisi}, {Ram{\'{\i}}rez}, {Mel{\'e}ndez}, {Bensby}, \&
  {Feltzing}}]{Casagrande2011}
{Casagrande}, L., {Sch{\"o}nrich}, R., {Asplund}, M., {et~al.} 2011, \aap, 530,
  A138

\bibitem[{{Casagrande} \& {VandenBerg}(2018)}]{Casagrande2018}
{Casagrande}, L., \& {VandenBerg}, D.~A. 2018, \mnras, 479, L102

\bibitem[{{Collier Cameron} \& {Jardine}(2018)}]{CollierCameron2018}
{Collier Cameron}, A., \& {Jardine}, M. 2018, \mnras, 476, 2542

\bibitem[{{Cumming} {et~al.}(2008){Cumming}, {Butler}, {Marcy}, {Vogt},
  {Wright}, \& {Fischer}}]{Cumming2008}
{Cumming}, A., {Butler}, R.~P., {Marcy}, G.~W., {et~al.} 2008, \pasp, 120, 531

\bibitem[{{Eggleton} {et~al.}(1998){Eggleton}, {Kiseleva}, \&
  {Hut}}]{Eggleton1998}
{Eggleton}, P.~P., {Kiseleva}, L.~G., \& {Hut}, P. 1998, \apj, 499, 853

\bibitem[{{Essick} \& {Weinberg}(2016)}]{Essick2016}
{Essick}, R., \& {Weinberg}, N.~N. 2016, \apj, 816, 18

\bibitem[{Evans(2018)}]{Evans2018}
Evans, D.~F. 2018, Research Notes of the {AAS}, 2, 20

\bibitem[{{Fischer} \& {Valenti}(2005)}]{Fischer2005}
{Fischer}, D.~A., \& {Valenti}, J. 2005, \apj, 622, 1102

\bibitem[{{Fressin} {et~al.}(2013){Fressin}, {Torres}, {Charbonneau}, {Bryson},
  {Christiansen}, {Dressing}, {Jenkins}, {Walkowicz}, \&
  {Batalha}}]{Fressin2013}
{Fressin}, F., {Torres}, G., {Charbonneau}, D., {et~al.} 2013, \apj, 766, 81

\bibitem[{{Gaia Collaboration} {et~al.}(2016){Gaia Collaboration}, {Prusti},
  {de Bruijne}, {Brown}, {Vallenari}, {Babusiaux}, {Bailer-Jones}, {Bastian},
  {Biermann}, {Evans}, \& et~al.}]{GaiaMission}
{Gaia Collaboration}, {Prusti}, T., {de Bruijne}, J.~H.~J., {et~al.} 2016,
  \aap, 595, A1

\bibitem[{{Gaia Collaboration} {et~al.}(2018{\natexlab{a}}){Gaia
  Collaboration}, {Babusiaux}, {van Leeuwen}, {Barstow}, {Jordi}, {Vallenari},
  {Bossini}, {Bressan}, {Cantat-Gaudin}, {van Leeuwen}, \& et~al.}]{GaiaDR2HR}
{Gaia Collaboration}, {Babusiaux}, C., {van Leeuwen}, F., {et~al.}
  2018{\natexlab{a}}, \aap, 616, A10

\bibitem[{{Gaia Collaboration} {et~al.}(2018{\natexlab{b}}){Gaia
  Collaboration}, {Brown}, {Vallenari}, {Prusti}, {de Bruijne}, {Babusiaux},
  {Bailer-Jones}, {Biermann}, {Evans}, {Eyer}, \& et~al.}]{GaiaDR2}
{Gaia Collaboration}, {Brown}, A.~G.~A., {Vallenari}, A., {et~al.}
  2018{\natexlab{b}}, \aap, 616, A1

\bibitem[{{Hansen}(2012)}]{Hansen2012}
{Hansen}, B.~M.~S. 2012, \apj, 757, 6

\bibitem[{{Hoyer} {et~al.}(2016){Hoyer}, {Pall{\'e}}, {Dragomir}, \&
  {Murgas}}]{Hoyer2016}
{Hoyer}, S., {Pall{\'e}}, E., {Dragomir}, D., \& {Murgas}, F. 2016, \aj, 151,
  137

\bibitem[{{Hsu} {et~al.}(2019){Hsu}, {Ford}, {Ragozzine}, \& {Ashby}}]{Hsu2019}
{Hsu}, D.~C., {Ford}, E.~B., {Ragozzine}, D., \& {Ashby}, K. 2019, arXiv
  e-prints, arXiv:1902.01417

\bibitem[{Husnoo {et~al.}(2012)Husnoo, Pont, Mazeh, Fabrycky, Hébrard, Bouchy,
  \& Shporer}]{Husnoo2012}
Husnoo, N., Pont, F., Mazeh, T., {et~al.} 2012, Monthly Notices of the Royal
  Astronomical Society, 422, 3151

\bibitem[{{Hut}(1981)}]{Hut1981}
{Hut}, P. 1981, \aap, 99, 126

\bibitem[{{Jackson} {et~al.}(2009){Jackson}, {Barnes}, \&
  {Greenberg}}]{Jackson2009}
{Jackson}, B., {Barnes}, R., \& {Greenberg}, R. 2009, \apj, 698, 1357

\bibitem[{{Jackson} {et~al.}(2008){Jackson}, {Greenberg}, \&
  {Barnes}}]{Jackson2008}
{Jackson}, B., {Greenberg}, R., \& {Barnes}, R. 2008, \apj, 678, 1396

\bibitem[{{Jiang} {et~al.}(2016){Jiang}, {Lai}, {Savushkin}, {Mkrtichian},
  {Antonyuk}, {Griv}, {Hsieh}, \& {Yeh}}]{Jiang2016}
{Jiang}, I.-G., {Lai}, C.-Y., {Savushkin}, A., {et~al.} 2016, \aj, 151, 17

\bibitem[{{Katz} {et~al.}(2019){Katz}, {Sartoretti}, {Cropper}, {Panuzzo},
  {Seabroke}, {Viala}, {Benson}, {Blomme}, {Jasniewicz}, \&
  {Jean-Antoine}}]{GaiaDR2RV}
{Katz}, D., {Sartoretti}, P., {Cropper}, M., {et~al.} 2019, \aap, 622, A205

\bibitem[{Labadie-Bartz {et~al.}(2019)Labadie-Bartz, Rodriguez, Stassun,
  Ciardi, Penev, Johnson, Gaudi, Col{\'{o}}n, Bieryla, Latham, Pepper, Collins,
  Evans, Relles, Siverd, Bento, Yao, Stockdale, Tan, Zhou, Eastman, Albrow,
  Bayliss, Beatty, Berlind, Bozza, Calkins, Cohen, Curtis, Esquerdo, Feliz,
  Fulton, Gregorio, James, Jensen, Johnson, Johnson, Joner, Kasper, Kielkopf,
  Kuhn, Lund, Malpas, Manner, McCrady, McLeod, Oberst, Penny, Reed, Sliski,
  Stephens, Stevens, Villanueva, Wittenmyer, Wright, \&
  Zambelli}]{LabadieBartz2019}
Labadie-Bartz, J., Rodriguez, J.~E., Stassun, K.~G., {et~al.} 2019, The
  Astrophysical Journal Supplement Series, 240, 13

\bibitem[{{Lai}(2012)}]{Lai2012}
{Lai}, D. 2012, \mnras, 423, 486

\bibitem[{{Levrard} {et~al.}(2009){Levrard}, {Winisdoerffer}, \&
  {Chabrier}}]{Levrard2009}
{Levrard}, B., {Winisdoerffer}, C., \& {Chabrier}, G. 2009, \apjl, 692, L9

\bibitem[{{Lindegren} {et~al.}(2018){Lindegren}, {Hern{\'a}ndez}, {Bombrun},
  {Klioner}, {Bastian}, {Ramos-Lerate}, {de Torres}, {Steidelm{\"u}ller},
  {Stephenson}, {Hobbs}, {Lammers}, {Biermann}, {Geyer}, {Hilger}, {Michalik},
  {Stampa}, {McMillan}, {Casta{\~n}eda}, {Clotet}, {Comoretto}, {Davidson},
  {Fabricius}, {Gracia}, {Hambly}, {Hutton}, {Mora}, {Portell}, {van Leeuwen},
  {Abbas}, {Abreu}, {Altmann}, {Andrei}, {Anglada}, {Balaguer-N{\'u}{\~n}ez},
  {Barache}, {Becciani}, {Bertone}, {Bianchi}, {Bouquillon}, {Bourda},
  {Br{\"u}semeister}, {Bucciarelli}, {Busonero}, {Buzzi}, {Cancelliere},
  {Carlucci}, {Charlot}, {Cheek}, {Crosta}, {Crowley}, {de Bruijne}, {de
  Felice}, {Drimmel}, {Esquej}, {Fienga}, {Fraile}, {Gai}, {Garralda},
  {Gonz{\'a}lez-Vidal}, {Guerra}, {Hauser}, {Hofmann}, {Holl}, {Jordan},
  {Lattanzi}, {Lenhardt}, {Liao}, {Licata}, {Lister}, {L{\"o}ffler},
  {Marchant}, {Martin-Fleitas}, {Messineo}, {Mignard}, {Morbidelli}, {Poggio},
  {Riva}, {Rowell}, {Salguero}, {Sarasso}, {Sciacca}, {Siddiqui}, {Smart},
  {Spagna}, {Steele}, {Taris}, {Torra}, {van Elteren}, {van Reeven}, \&
  {Vecchiato}}]{Lindegren2018}
{Lindegren}, L., {Hern{\'a}ndez}, J., {Bombrun}, A., {et~al.} 2018, \aap, 616,
  A2

\bibitem[{{Maciejewski} {et~al.}(2016){Maciejewski}, {Dimitrov},
  {Fern{\'a}ndez}, {Sota}, {Nowak}, {Ohlert}, {Nikolov}, {Bukowiecki}, {Hinse},
  {Pall{\'e}}, {Tingley}, {Kjurkchieva}, {Lee}, \& {Lee}}]{Maciejewski2016}
{Maciejewski}, G., {Dimitrov}, D., {Fern{\'a}ndez}, M., {et~al.} 2016, \aap,
  588, L6

\bibitem[{{Maciejewski} {et~al.}(2018){Maciejewski}, {Fern{\'a}ndez},
  {Aceituno}, {Mart{\'{\i}}n-Ruiz}, {Ohlert}, {Dimitrov}, {Szyszka}, {von
  Essen}, {Mugrauer}, {Bischoff}, {Michel}, {Mallonn}, {Stangret}, \&
  {Mo{\'z}dzierski}}]{Maciejewski2018}
{Maciejewski}, G., {Fern{\'a}ndez}, M., {Aceituno}, F., {et~al.} 2018, \actaa,
  68, 371

\bibitem[{{Marchetti} {et~al.}(2018){Marchetti}, {Rossi}, \&
  {Brown}}]{Marchetti2018}
{Marchetti}, T., {Rossi}, E.~M., \& {Brown}, A.~G.~A. 2018, \mnras,
  arXiv:1804.10607

\bibitem[{{Matsumura} {et~al.}(2010){Matsumura}, {Peale}, \&
  {Rasio}}]{Matsumura2010}
{Matsumura}, S., {Peale}, S.~J., \& {Rasio}, F.~A. 2010, \apj, 725, 1995

\bibitem[{{Maxted} {et~al.}(2015){Maxted}, {Serenelli}, \&
  {Southworth}}]{Maxted2015}
{Maxted}, P.~F.~L., {Serenelli}, A.~M., \& {Southworth}, J. 2015, \aap, 577,
  A90

\bibitem[{{Mayor} \& {Queloz}(1995)}]{Mayor1995}
{Mayor}, M., \& {Queloz}, D. 1995, \nat, 378, 355

\bibitem[{McKinney {et~al.}(2010)}]{pandas}
McKinney, W., {et~al.} 2010, in Proceedings of the 9th Python in Science
  Conference, Vol. 445, Austin, TX, 51--56

\bibitem[{{McQuillan} {et~al.}(2013){McQuillan}, {Mazeh}, \&
  {Aigrain}}]{McQuillan2013}
{McQuillan}, A., {Mazeh}, T., \& {Aigrain}, S. 2013, \apjl, 775, L11

\bibitem[{{McTier} \& {Kipping}(2019)}]{McTier2019}
{McTier}, M., \& {Kipping}, D. 2019, arXiv e-prints, arXiv:1906.02663

\bibitem[{{Meibom} \& {Mathieu}(2005)}]{Meibom2005}
{Meibom}, S., \& {Mathieu}, R.~D. 2005, \apj, 620, 970

\bibitem[{{Milliman} {et~al.}(2014){Milliman}, {Mathieu}, {Geller}, {Gosnell},
  {Meibom}, \& {Platais}}]{Milliman2014}
{Milliman}, K.~E., {Mathieu}, R.~D., {Geller}, A.~M., {et~al.} 2014, \aj, 148,
  38

\bibitem[{{Ogilvie}(2009)}]{Ogilvie2009}
{Ogilvie}, G.~I. 2009, \mnras, 396, 794

\bibitem[{{Ogilvie}(2014)}]{Ogilvie2014}
---. 2014, \araa, 52, 171

\bibitem[{{Ogilvie} \& {Lin}(2004)}]{Ogilvie2004}
{Ogilvie}, G.~I., \& {Lin}, D.~N.~C. 2004, \apj, 610, 477

\bibitem[{{Patra} {et~al.}(2017){Patra}, {Winn}, {Holman}, {Yu}, {Deming}, \&
  {Dai}}]{Patra2017}
{Patra}, K.~C., {Winn}, J.~N., {Holman}, M.~J., {et~al.} 2017, \aj, 154, 4

\bibitem[{{Pecaut} \& {Mamajek}(2013)}]{Mamajek2013}
{Pecaut}, M.~J., \& {Mamajek}, E.~E. 2013, \apjs, 208, 9

\bibitem[{{Penev} {et~al.}(2018){Penev}, {Bouma}, {Winn}, \&
  {Hartman}}]{Penev2018}
{Penev}, K., {Bouma}, L.~G., {Winn}, J.~N., \& {Hartman}, J.~D. 2018, \aj, 155,
  165

\bibitem[{{Penev} {et~al.}(2012){Penev}, {Jackson}, {Spada}, \&
  {Thom}}]{Penev2012}
{Penev}, K., {Jackson}, B., {Spada}, F., \& {Thom}, N. 2012, \apj, 751, 96

\bibitem[{{Penev} \& {Sasselov}(2011)}]{Penev2011}
{Penev}, K., \& {Sasselov}, D. 2011, \apj, 731, 67

\bibitem[{{Pont}(2009)}]{Pont2009}
{Pont}, F. 2009, \mnras, 396, 1789

\bibitem[{Price-Whelan(2018)}]{PriceWhelan2018}
Price-Whelan, A. 2018, doi:10.5281/zenodo.1228136

\bibitem[{{Price-Whelan} {et~al.}(2018){Price-Whelan}, {Sip{\H{o}}cz},
  {G{\"u}nther}, {Lim}, {Crawford}, {Conseil}, {Shupe}, {Craig}, {Dencheva},
  {Ginsburg}, {VanderPlas}, {Bradley}, {P{\'e}rez-Su{\'a}rez}, {de Val-Borro},
  {Paper Contributors}, {Aldcroft}, {Cruz}, {Robitaille}, {Tollerud},
  {Coordination Committee}, {Ardelean}, {Babej}, {Bach}, {Bachetti}, {Bakanov},
  {Bamford}, {Barentsen}, {Barmby}, {Baumbach}, {Berry}, {Biscani}, {Boquien},
  {Bostroem}, {Bouma}, {Brammer}, {Bray}, {Breytenbach}, {Buddelmeijer},
  {Burke}, {Calderone}, {Cano Rodr{\'\i}guez}, {Cara}, {Cardoso}, {Cheedella},
  {Copin}, {Corrales}, {Crichton}, {D{\textquoteright}Avella}, {Deil},
  {Depagne}, {Dietrich}, {Donath}, {Droettboom}, {Earl}, {Erben}, {Fabbro},
  {Ferreira}, {Finethy}, {Fox}, {Garrison}, {Gibbons}, {Goldstein}, {Gommers},
  {Greco}, {Greenfield}, {Groener}, {Grollier}, {Hagen}, {Hirst}, {Homeier},
  {Horton}, {Hosseinzadeh}, {Hu}, {Hunkeler}, {Ivezi{\'c}}, {Jain}, {Jenness},
  {Kanarek}, {Kendrew}, {Kern}, {Kerzendorf}, {Khvalko}, {King}, {Kirkby},
  {Kulkarni}, {Kumar}, {Lee}, {Lenz}, {Littlefair}, {Ma}, {Macleod},
  {Mastropietro}, {McCully}, {Montagnac}, {Morris}, {Mueller}, {Mumford},
  {Muna}, {Murphy}, {Nelson}, {Nguyen}, {Ninan}, {N{\"o}the}, {Ogaz}, {Oh},
  {Parejko}, {Parley}, {Pascual}, {Patil}, {Patil}, {Plunkett}, {Prochaska},
  {Rastogi}, {Reddy Janga}, {Sabater}, {Sakurikar}, {Seifert}, {Sherbert},
  {Sherwood-Taylor}, {Shih}, {Sick}, {Silbiger}, {Singanamalla}, {Singer},
  {Sladen}, {Sooley}, {Sornarajah}, {Streicher}, {Teuben}, {Thomas},
  {Tremblay}, {Turner}, {Terr{\'o}n}, {van Kerkwijk}, {de la Vega}, {Watkins},
  {Weaver}, {Whitmore}, {Woillez}, {Zabalza}, \& {Contributors}}]{astropy:2018}
{Price-Whelan}, A.~M., {Sip{\H{o}}cz}, B.~M., {G{\"u}nther}, H.~M., {et~al.}
  2018, \aj, 156, 123

\bibitem[{{Rasio} \& {Ford}(1996)}]{Rasio1996b}
{Rasio}, F.~A., \& {Ford}, E.~B. 1996, Science, 274, 954

\bibitem[{{Rasio} {et~al.}(1996){Rasio}, {Tout}, {Lubow}, \&
  {Livio}}]{Rasio1996}
{Rasio}, F.~A., {Tout}, C.~A., {Lubow}, S.~H., \& {Livio}, M. 1996, \apj, 470,
  1187

\bibitem[{{Sanchis-Ojeda} {et~al.}(2014){Sanchis-Ojeda}, {Rappaport}, {Winn},
  {Kotson}, {Levine}, \& {El Mellah}}]{SanchisOjeda2014}
{Sanchis-Ojeda}, R., {Rappaport}, S., {Winn}, J.~N., {et~al.} 2014, \apj, 787,
  47

\bibitem[{{Santerne} {et~al.}(2016){Santerne}, {Moutou}, {Tsantaki}, {Bouchy},
  {H{\'e}brard}, {Adibekyan}, {Almenara}, {Amard}, {Barros}, {Boisse},
  {Bonomo}, {Bruno}, {Courcol}, {Deleuil}, {Demangeon}, {D{\'{\i}}az},
  {Guillot}, {Havel}, {Montagnier}, {Rajpurohit}, {Rey}, \&
  {Santos}}]{Santerne2016}
{Santerne}, A., {Moutou}, C., {Tsantaki}, M., {et~al.} 2016, \aap, 587, A64

\bibitem[{{Santos} {et~al.}(2004){Santos}, {Israelian}, \&
  {Mayor}}]{Santos2004}
{Santos}, N.~C., {Israelian}, G., \& {Mayor}, M. 2004, \aap, 415, 1153

\bibitem[{{Santos} {et~al.}(2013){Santos}, {Sousa}, {Mortier}, {Neves},
  {Adibekyan}, {Tsantaki}, {Delgado Mena}, {Bonfils}, {Israelian}, {Mayor}, \&
  {Udry}}]{Santos2013}
{Santos}, N.~C., {Sousa}, S.~G., {Mortier}, A., {et~al.} 2013, \aap, 556, A150

\bibitem[{{Schlaufman}(2010)}]{Schlaufman2010}
{Schlaufman}, K.~C. 2010, \apj, 719, 602

\bibitem[{{Schlaufman} \& {Winn}(2013)}]{Schlaufman2013}
{Schlaufman}, K.~C., \& {Winn}, J.~N. 2013, \apj, 772, 143

\bibitem[{{Silva Aguirre} {et~al.}(2018){Silva Aguirre}, {Bojsen-Hansen},
  {Slumstrup}, {Casagrande}, {Kawata}, {Ciuc{\v a}}, {Handberg}, {Lund},
  {Mosumgaard}, {Huber}, {Johnson}, {Pinsonneault}, {Serenelli}, {Stello},
  {Tayar}, {Bird}, {Cassisi}, {Hon}, {Martig}, {Nissen}, {Rix},
  {Sch{\"o}nrich}, {Sahlholdt}, {Trick}, \& {Yu}}]{SilvaAguirre2018}
{Silva Aguirre}, V., {Bojsen-Hansen}, M., {Slumstrup}, D., {et~al.} 2018,
  \mnras, 475, 5487

\bibitem[{{Soderblom}(2010)}]{Soderblom2010}
{Soderblom}, D.~R. 2010, \araa, 48, 581

\bibitem[{{Sousa} {et~al.}(2018){Sousa}, {Adibekyan}, {Delgado-Mena}, {Santos},
  {Andreasen}, {Ferreira}, {Tsantaki}, {Barros}, {Demangeon}, {Israelian},
  {Faria}, {Figueira}, {Mortier}, {Brand{\~a}o}, {Montalto}, {Rojas-Ayala}, \&
  {Santerne}}]{Sousa2018}
{Sousa}, S.~G., {Adibekyan}, V., {Delgado-Mena}, E., {et~al.} 2018, \aap, 620,
  A58

\bibitem[{{Teitler} \& {K{\"o}nigl}(2014)}]{Teitler2014}
{Teitler}, S., \& {K{\"o}nigl}, A. 2014, \apj, 786, 139

\bibitem[{Torres {et~al.}(2010)Torres, Andersen, \& Gim{\'e}nez}]{Torres2010}
Torres, G., Andersen, J., \& Gim{\'e}nez, A. 2010, The Astronomy and
  Astrophysics Review, 18, 67

\bibitem[{{Udry} \& {Santos}(2007)}]{Udry2007}
{Udry}, S., \& {Santos}, N.~C. 2007, \araa, 45, 397

\bibitem[{{Wenger} {et~al.}(2000){Wenger}, {Ochsenbein}, {Egret}, {Dubois},
  {Bonnarel}, {Borde}, {Genova}, {Jasniewicz}, {Lalo{\"e}}, {Lesteven}, \&
  {Monier}}]{SIMBAD}
{Wenger}, M., {Ochsenbein}, F., {Egret}, D., {et~al.} 2000, \aaps, 143, 9

\bibitem[{{Wilkins} {et~al.}(2017){Wilkins}, {Delrez}, {Barker}, {Deming},
  {Hamilton}, {Gillon}, \& {Jehin}}]{Wilkins2017}
{Wilkins}, A.~N., {Delrez}, L., {Barker}, A.~J., {et~al.} 2017, \apjl, 836, L24

\bibitem[{{Winn} {et~al.}(2010){Winn}, {Fabrycky}, {Albrecht}, \&
  {Johnson}}]{Winn2010}
{Winn}, J.~N., {Fabrycky}, D., {Albrecht}, S., \& {Johnson}, J.~A. 2010, \apj,
  718, L145

\bibitem[{{Wright} {et~al.}(2012){Wright}, {Marcy}, {Howard}, {Johnson},
  {Morton}, \& {Fischer}}]{Wright2012}
{Wright}, J.~T., {Marcy}, G.~W., {Howard}, A.~W., {et~al.} 2012, \apj, 753, 160

\bibitem[{{Zahn}(1977)}]{Zahn1977}
{Zahn}, J.-P. 1977, \aap, 57, 383

\bibitem[{{Zhou} {et~al.}(2019){Zhou}, {Huang}, {Bakos}, {Hartman}, {Latham},
  {Quinn}, {Collins}, {Winn}, {Kovacs}, {Csubry}, {Bhatti}, {Penev}, {Bieryla},
  {Esquerdo}, {Berlind}, {Calkins}, {de Val-Borro}, {Noyes}, {L{\'a}z{\'a}r},
  {Papp}, {Sari}, {Kovacs}, {Buchhave}, {Szklen{\'a}r}, {Beky}, {Johnson},
  {Stassun}, {Shporer}, {Wong}, {Espinoza}, {Bayliss}, {Howell}, {Hellier},
  {Anderson}, {West}, {Brown}, {Schanche}, {Barkaoui}, {Pozuelos}, {Gillon},
  {Jehin}, {Benkhaldoun}, {Daassou}, {Ricker}, {Vanderspek}, {Seager},
  {Jenkins}, {Lissauer}, {Collins}, {Gan}, {Hart}, {Horne}, {Kielkopf},
  {Nielsen}, {Nishiumi}, {Narita}, {Palle}, {Relles}, {Sefako}, {Tan},
  {Davies}, {Goeke}, {Guerrero}, {Haworth}, \& {Villanueva}}]{Zhou2019}
{Zhou}, G., {Huang}, C., {Bakos}, G., {et~al.} 2019, arXiv e-prints,
  arXiv:1906.00462

\end{thebibliography}

\appendix 
\label{app1}
\noindent \citet{Lindegren2018} and \cite{Marchetti2018} suggest the following quality cuts to ensure reliable astrometry.  We apply them to our field star sample. Cuts 4 and 8 are cuts C.1 and C.2 of \citet{Lindegren2018}, where $u$ is the unit weight error and $E$ is the $\texttt{phot\_bp\_rp\_excess\_factor}$.  Both cuts are related to problems that arise due to crowding.  Cut C.1 removes sources for which the single-star parallax model does not fit well, as two nearby objects are instead mistaken for one object with a large parallax.  Cut C.2 removes faint objects in crowded regions, for which there are significant photometric errors in the $G_{BP}$ and $G_{RP}$ magnitudes.  We also impose astrometric quality cuts 1-4 to the hot Jupiter host sample.  We do not apply cut 6 to the hot Jupiter host sample because it is known that the reflex motion of giant planet host stars can result in excess noise in the astrometric fitting \citep{Evans2018}.  We do not apply cut 7 to the hot Jupiter host sample because the hot Jupiter host radial velocities come from ground-based radial velocities.  We apply cuts 1, 5, and 8-10 to the Pleiades members confirmed by Gaia-DR2 when generating the main sequence $(G_{BP}-G_{RP})$--$M_G$ polynomial fit.  Overall, these cuts are designed to produce a sample with high-quality astrometry.
\begin{enumerate}
    \item $\texttt{parallax\_over\_error} > 10$
    \item $-0.23 < \texttt{mean\_varpi\_factor} < 0.36$
    \item $\texttt{visibility\_periods\_used} > 8$
    \item $u < 1.2*\texttt{MAX}(1,\exp{(-0.2*\texttt{phot\_g\_mean\_mag}-19.5}))$
    \item $\texttt{astrometric\_gof\_al} <3$
    \item $\texttt{astrometric\_excess\_noise\_sig}<2$
    \item $\texttt{rv\_nb\_transits}>5$
    \item $1.0+0.0015*\texttt{bp\_rp}^2 < E < 1.3+0.06*\texttt{bp\_rp}^2$
    \item $\texttt{phot\_bp\_mean\_flux\_over\_error} >10$
    \item $\texttt{phot\_rp\_mean\_flux\_over\_error} >10$
\end{enumerate}

\end{document}